\documentclass[twocolumn]{aastex63}
\usepackage{graphicx} 
\usepackage{hyperref}
\usepackage{xcolor}

\usepackage{lineno}

\newcommand{\codename}[1]{\texttt{#1}}
\newcommand{\korg}{\codename{Korg}}

\newcommand{\revision}[1]{#1}

\newcommand{\apogee}{APOGEE}
\newcommand{\breta}{$\mathrm{Br\text{--}\eta}$}
\newcommand{\brlamda}{$\mathrm{Br\text{--}\lambda}$}
\newcommand{\halpha}{$\mathrm{H}\text{--}\alpha$}

\newcommand\textmicrometer{~{\textmu}m}
\newcommand\micrometer{%
\ifmmode\textnormal{\textmicrometer}%
\else\textmicrometer%
\fi}

\begin{document}
\title{Korg: fitting, model atmosphere interpolation, and Brackett lines}
\shorttitle{Korg: fitting, model atmosphere interpolation, and Brackett lines}

\author[0000-0001-7339-5136]{Adam J. Wheeler}
\affiliation{Department of Astronomy, Ohio State University, McPherson Laboratory, 140 West 18th Avenue, Columbus, Ohio, USA}
\author[0000-0003-0174-0564]{Andrew R. Casey}
\affiliation{School of Physics \& Astronomy, Monash University, Victoria, Australia}
\affiliation{Center of Excellence for Astrophysics in Three Dimensions (ASTRO-3D)}
\author[0000-0002-7918-3086]{Matthew W. Abruzzo}
\affiliation{Department of Physics and Astronomy, University of Pittsburgh, 100 Allen Hall 3941 O’Hara Street, Pittsburgh, PA 15260 USA}

\correspondingauthor{Adam Wheeler}
\email{wheeler.883@osu.edu}

\begin{abstract}
\noindent
We describe several updates to \korg{}, a package for 1D LTE spectral synthesis of FGKM stars.
Built-in functions to fit observed spectra via synthesis or equivalent widths make it easy to take advantage of \korg{}'s automatic differentiation.
Comparison to a past analysis of 18 Sco shows that we obtain significantly reduced line-to-line abundance scatter with \korg{}.
Fitting and synthesis are facilitated by a rigorously-tested model atmosphere interpolation method, which introduces negligible error to synthesized spectra for stars with $T_\mathrm{eff} \gtrsim 4000\,\mathrm{K}$.
For cooler stars, atmosphere interpolation is complicated by the presence of molecules, though we demonstrate an adequate method for cool dwarfs.
The chemical equilibrium solver has been extended to include polyatomic and charged molecules, extending \korg{}'s regime of applicability to M stars.
We also discuss a common oversight regarding the synthesis of hydrogen lines in the infrared, and show that \korg{}'s Brackett line profiles are a much closer match to observations than others available.
Documentation, installation instructions, and tutorials are available at \url{https://github.com/ajwheeler/Korg.jl}.  

\end{abstract}


\section{Introduction}
\korg{}, first presented in \citet{wheelerKORGModern1D2023}, is a stellar spectral synthesis code for computing spectra of FGKM stars.
It it been written with both flexibility and performance in mind, and is faster than similar codes by a factor of 1-100.
It's central assumptions are of 1D geometry and local thermodynamic equilibrium, which, while not state-of-the art theoretically, are well understood and significantly speed up and simplify the calculation of spectra.
\korg{} is written in pure Julia, which makes it easy to use in scripts or an interactive programming environment from either Julia or Python.
It also supports automatic differentiation, which can speed up calculation involving derivatives by an order of magnitude or more.

This paper describes several updates to \korg{} which aim to make it useful to researchers in more contexts
(\citealp{wheelerKORGModern1D2023} provides an overview of the code, here we focus on changes only).
Section \ref{sec:fit} discusses new routines for automatic fitting of observed data via synthesis or equivalent widths, as well as the employed model atmosphere interpolation scheme.
This requires the interpolation of model atmospheres, which is not often given the attention it requires.
Section \ref{sec:internal} discusses several model improvements, most notably a more sophisticated treatment of molecules (section \ref{sec:eos}), which allows for synthesis of M star spectra.
Section \ref{sec:conclusions} summarizes our conclusions.

\section{Fitting spectra via direct comparison or equivalent widths }\label{sec:fit}
\korg{} now includes functions for inferring parameters and abundances via direct fitting of the observed spectrum (synthesis) or via equivalent widths, \texttt{find\_best\_fit\_params} and \texttt{ews\_to\_abundances}.

\subsection{Fitting via synthesis}
The \texttt{find\_best\_fit\_params} function is designed to make it as easy as possible to take advantage of \korg{}'s automatic differentiation capabilities, even when called from Python.
It takes as input an observed rectified spectrum with known error, a linelist and initial guesses for each parameter to be fit.
Fitting is performed with the Broyden–Fletcher–Goldfarb–Shanno (BFGS) algorithm (see, e.g. \citealp{nocedalNumericalOptimizationSpringer2006}) as implemented in the \texttt{Optim.jl} package \citep{mogensenJuliaNLSolversNLsolveJl2020}.
The fit is performed within wavelength windows specified by the user, with minimal computational overhead imposed by non-contiguous windows.
Any subset of the following parameters can be fit, with the remainder fixed to default or user-specified values:
\begin{itemize}
    \item $T_\mathrm{eff}$: the effective temperature.  
    \item $\log g$: the surface gravity.
    \item \texttt{m\_H}: the metallicity of the star, in dex. (defaults to $0.0$).
    \item \revision{\texttt{alpha\_H}: the $\alpha$-element abundance of the star, in dex. (defaults to the value of \texttt{m\_H}.) Here, the $\alpha$-elements are interpreted as those with even atomic numbers from C to Ti.}
    \item Individual element abundances, e.g. `Na`, specify the solar-relative ([X/H]) abundance of that element (defaults to the value of the metallicity) .
    \item \texttt{vmic}: the microturbulent velocity, in km/s (defaults to $1.0$).
    \item \texttt{vsini}: the projected rotational velocity of the star, in km/s (defaults to  $0.0$).
    \item \texttt{epsilon}: the linear limb-darkening coefficient (defaults to $0.6$). This is used only for applying rotational broadening.  
\end{itemize}
By default, convolution with the line-spread function (LSF) is automatically handled via construction of a sparse matrix, but user-specified LSFs can be substituted if desired.

\subsection{Fitting via equivalent widths}
The \texttt{ews\_to\_abundances} function computes abundances of individual atomic transitions, given some measured equivalent widths and a model atmosphere.
This functionality is inspired by the widely used  \texttt{ABFIND} routine in \texttt{MOOG} \citep{snedenNitrogenAbundanceVery1973}, but differs in its approach.
While \texttt{MOOG} computes a fictitious curve of growth for each species, \korg{} synthesizes each line in wavelength segments and integrates the rectified profile to compute a model equivalent width.
Nearby lines are segmented into groups to ensure wavelength segments do not overlap, and that absorption from nearby lines do not contribute.
Chemical equilibrium and continuum opacity calculations are not duplicated within groups of wavelength segments to avoid redundant computation.

Individual line abundances are computed by first synthesizing each line at an initial abundance\footnote{Abundances are in the standard absolute format used by spectroscopists, $A(X) = \log_{10} \frac{n_X}{n_\mathrm{H}} + 12$.} specified by the user, $[A(X)]_0$, to obtain an equivalent width, $W_0$.
Lines are assumed to be on the linear part of the curve of grown, so then,
\begin{equation}
    A(X) = [A(X)]_0 + \log_{10} \frac{W}{W_0} \quad ,
\end{equation}
where $W$ is the observed equivalent width and $A(X)$ is the resulting abundance.
While this procedure could be sped up by using a fictitious curve of growth for each line, synthesizing each line once ensures the most accurate calculation possible.

\begin{figure*} 
    \centering
    \includegraphics[width=\textwidth]{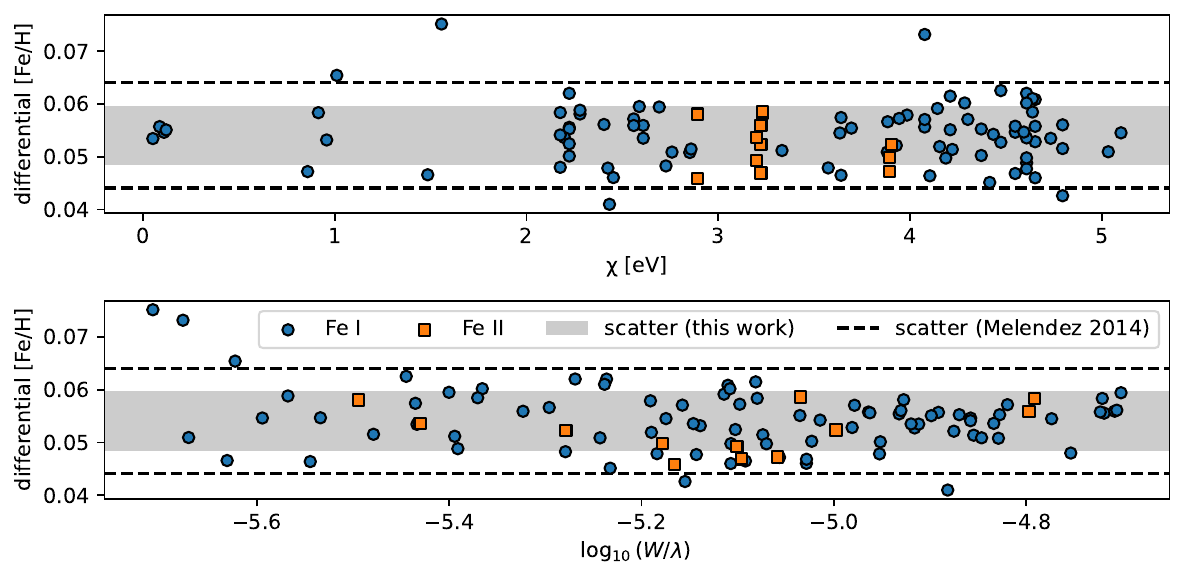}
    \caption{Differential [Fe/H] abundances of 18 Sco (relative to the Sun) computed by \korg{} with equivalent widths and stellar parameters from \citet{melendez18ScoSolar2014}. The mean abundance is the same found by \citet{melendez18ScoSolar2014}, and the scatter in line abundances from \korg{} is nearly half that of \citet{melendez18ScoSolar2014} (0.0056 dex compared to 0.010 dex).
    }
    \label{fig:melendez}
\end{figure*}

Figure~\ref{fig:melendez} shows an example use of \texttt{ews\_to\_abundances} functionality.
Equivalent widths are measured and reported by \citet{melendez18ScoSolar2014} for the Sun and 18 Sco, a solar-like twin.
\korg{} was used to interpolate a model atmosphere for the Sun ($T_\mathrm{eff} = 5777\,\mathrm{K}$, $\log{g} = 4.44$, [Fe/H] = 0.0) and for 18 Sco, using the parameters reported by \citet{melendez18ScoSolar2014}: $T_\mathrm{eff} = 5823\,\mathrm{K}$, $\log{g} = 4.45$, $[\mathrm{Fe/H}] = 0.054$.
Because \korg{}'s model atmosphere grid (discussed in section \ref{sec:itp}) does not include $v_\mathrm{mic}$, both model atmospheres were interpolated with the default value for dwarfs, $v_\mathrm{mic} = 1 \,\mathrm{km\,s}^{-1}$.  
Subsequent syntheses of each profile used $v_\mathrm{mic} = 1 \,\mathrm{km\,s}^{-1}$ for the sun, and $v_\mathrm{mic} = 1.02\,\mathrm{km\,s}^{-1}$ for 18 Sco (the same as in \citealp{melendez18ScoSolar2014}).
The differential (18 Sco - Sun) abundances for neutral (circle) and singly ionized (square) transitions are shown as a function of excitation potential and reduced equivalent width.

\begin{figure}
    \centering
    \includegraphics[width=0.45\textwidth]{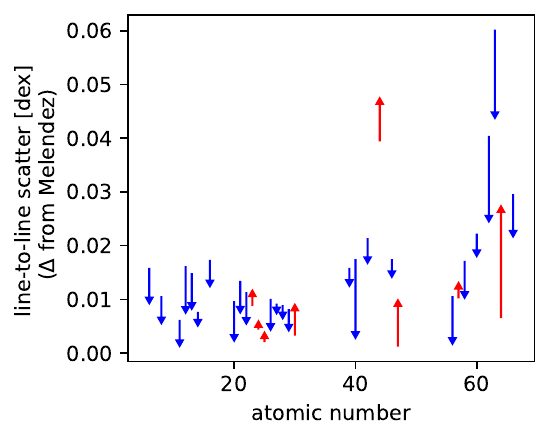}
    \caption{The line-to-line scatter in the abundance of each element measured using equivalent widths from \citet{melendez18ScoSolar2014}.  Comparison between \korg{} reanalysis and original abundances. Computing abundances with \korg{} results in significantly smaller scatter for most elements.}
    \label{fig:melendez_all}
\end{figure}

While the same stellar parameters and equivalent widths are used, the scatter in \korg{} line abundances is about half ($\sigma = 0.0056$ compared to $\sigma = 0.010$) that reported in \citet{melendez18ScoSolar2014}.
The weakest lines show the largest abundance differences between 18 Sco and the Sun, where even the few significant digits used to record the equivalent width (e.g., rounding from 12.15\,m\AA\ to 12.2\,m\AA) can translate to a systematic shift of 0.01\,dex for that line abundance.
\revision{Figure \ref{fig:melendez_all} compares the line-to-line scatter for all abundances measured using equivalent widths in \citet{melendez18ScoSolar2014}, demonstrating that the improved precision is not limited to iron lines.}
Excluding the spectral synthesis code used, the other notable differences between these analyses are the grid of model atmospheres (\citealp{melendez18ScoSolar2014} used Kurucz models), how they are interpolated (section~\ref{sec:itp}), and synthesizing lines instead of using a fictitious curve of growth.
Thus, \korg{}'s \texttt{ew\_to\_abundances} functionality is suitable for classical spectroscopic determination of stellar parameters, including high-precision differential-abundance research.

\subsection{Model atmosphere interpolation} \label{sec:itp}
In order to synthesize a spectrum for a star with arbitrary parameters, or to fit a observed spectrum, the input model atmosphere must be either constructed \emph{ad hoc} or interpolated to precise parameter values.
\korg{} now includes functionality to interpolate the Sloan Digital Sky Survey (SDSS) model atmosphere grid\footnote{\url{https://dr17.sdss.org/sas/dr17/apogee/spectro/speclib/atmos/marcs/MARCS_v3_2016/Readme_MARCS_v3_2016.txt}}, an expanded version of the MARCS \citep{gustafssonGridMARCSModel2008} grid documented in \citet{meszarosNewATLAS9MARCS2012}.
The grid contains 579,150 model atmospheres, which vary over five parameters: $T_\mathrm{eff}$ (2500 to 8000 K), $\log g$ (-0.5 to 5.5), [metals/H] (-2.5 to 1.0), [$\alpha$/Fe] (-1 to 1), and [C/Fe] (-1 to 1).
Atmospheres for some parameters (mostly in unphysical regimes, but also the tip of the red-giant branch) did not converge.
\citet{meszarosNewATLAS9MARCS2012} filled these in with radial basis function interpolation.
The atmospheres in the grid are divided into two groups.  
Those with $\log g > 3$ are in plane-parallel geometry and were computed with a microturbulent velocity of $1 \,\mathrm{km\,s}^{-1}$.
Those with $\log g < 3$ are in spherical geometry (with an assumed mass of $1 M_\odot$) and were computed with a microturbulent velocity of $2\,\mathrm{km\,s}^{-1}$.

When \korg{} takes model atmosphere as input, it uses five quantities defined at each layer of the atmosphere: temperature, number density, electron number density (used only as a starting guess, see section \ref{sec:eos}), physical depth, and the optical depth at 5000 \AA.
We experimented with various transforms of the parameters and found that better accuracy is obtained when number density and electron number density are $\log$ transformed, and when geometric depth is transformed with $\sinh^{-1}$.
\korg{} interpolates the transformed quantities as a function of the parameters over which the grid varies using simple multilinear interpolation (provided by the \codename{Interpolations.jl} package\footnote{\url{https://github.com/JuliaMath/Interpolations.jl}}). See the discussion below regarding cubic interpolation of model atmospheres for cool stars.

\begin{figure*}
    \centering
    \includegraphics[width=\textwidth]{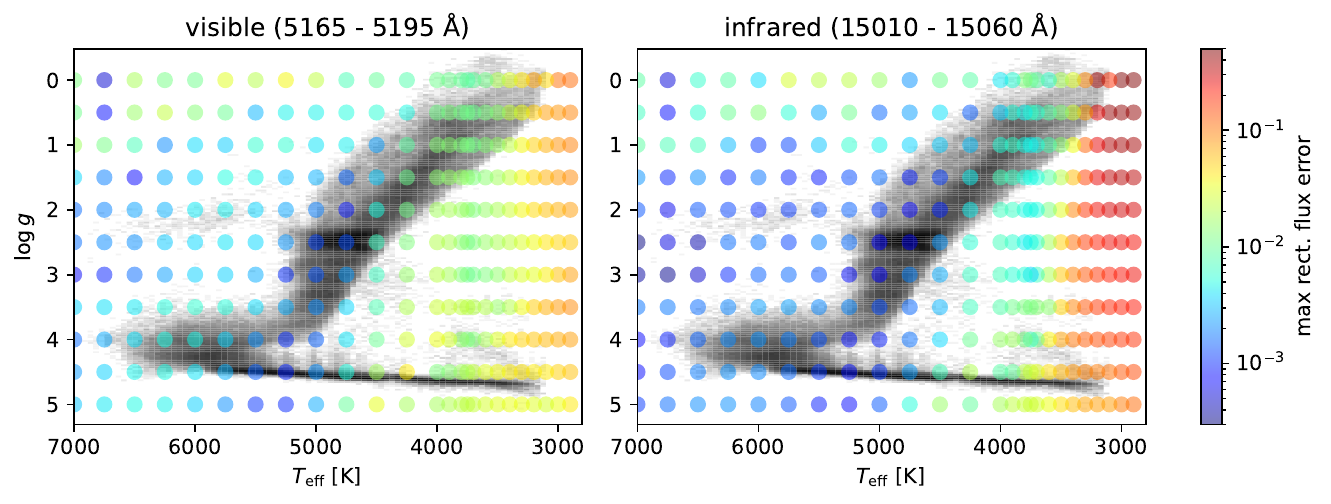}
    \caption{Estimated error introduced by interpolation as a function of $T_\mathrm{eff}$ and $\log g$ for solar abundances. The color of each point shows the largest flux deviation at any wavelength for spectra synthesized with the APOGEE DR 17 linelist \citep{abdurroufSeventeenthDataRelease2022} and convolved to $R = 22,500$ (the approximate APOGEE resolving power).  We note, however, that this plot is not sensitive to wavelength range, linelist, or $R$.  For reference, the distribution of stars in APOGEE DR17 is plotted in the background. The colorbar has been truncated for clarity; errors are smaller than $10^-6$ for some stellar parameters.}
    \label{fig:itp_err}
\end{figure*}

We use a simple procedure to quantify errors introduced by model atmosphere interpolation. 
First we construct a new, temporary grid by interpolating halfway between nodes in all dimensions,  then we interpolate back to the the original grid points.
Differences between the rectified spectra produced from a doubly interpolated atmosphere and the corresponding original atmosphere provides an upper bound on the error introduced by interpolation.

Figure \ref{fig:itp_err} shows the largest absolute error introduced to syntheses in the visible and infrared by the interpolation of model atmospheres in all five parameters.
Interpolations errors are slightly lower in the infrared than in the visible because, as noted in \citet{westendorpplazaHighprecisionInterpolationStellar2023}, traditional interpolation methods perform the worst for the highest and lowest atmospheric layers, and lines in the infrared are less sensitive to the upper atmosphere.
For most parts of the Kiel diagram, the error introduced by interpolation is negligible, but it becomes important for cool stars.
Because we are comparing rectified spectra, the largest errors are in the line cores, though for cool line-blanketed stars, the interpolation error manifests as a near-constant offset.
In contrast with \citet{meszarosInterpolationModelAtmospheres2013}, we find that there is no benefit to performing interpolation at the spectral level, rather than the atmospheric level (which is more costly unless spectra are precomputed).

\begin{figure}
    \centering
    \includegraphics[width=0.45 \textwidth]{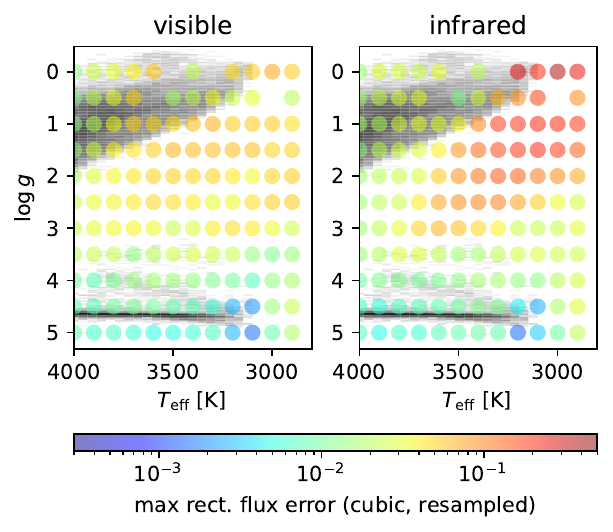}
    \caption{Estimated error introduced by interpolation using the method tailored for cool dwarfs, described in text. Changes which improved interpolation for cool dwarfs make it worse for cool giants. (Compare the main sequence and red-giant branch to those in Figure \ref{fig:itp_err}.)}
    \label{fig:cubic_itp_err}
\end{figure}

While model atmosphere interpolation works well for most stars, it is problematic for those with $T_\mathrm{eff} \lesssim 4000~\mathrm{K}$.
As an experiment, we developed an alternative interpolation method (not yet available in \codename{Korg}) tailored for cool dwarfs.
We found that $\tau_{5000}$, the optical depth at $5000~\mathrm{\AA}$, is particularly difficult to accurately interpolate in the cool-dwarf regime.
When using the default radiative transfer scheme, \korg{} uses $\tau_{5000}$ to anchor its calculations (see \citealp{wheelerKORGModern1D2023}).
The method tailored for cool dwarfs performs interpolation on atmospheres that have been resampled (via cubic interpolation) so that atmospheric depth is now indexed in terms of a shared set of fixed $\tau_{5000}$.
In other words, we no longer interpolate $\tau_{5000}$ between grid points.
For cool dwarfs, resampling the atmospheres had negligible effect on synthesized spectra at grid points and significantly reduced the interpolation error (estimated as described above).
Additionally, we interpolated between the resampled model atmospheres using cubic (rather than linear) interpolation.
The results of this procedure are shown in Figure \ref{fig:cubic_itp_err}.
Both resampling and cubic interpolation improve the interpolation error for cool dwarfs, reducing the error to the subpercent level.
Unfortunately, for cool giants, each change made the interpolation error worse.

\begin{figure*}
    \centering
    \includegraphics[width=\textwidth]{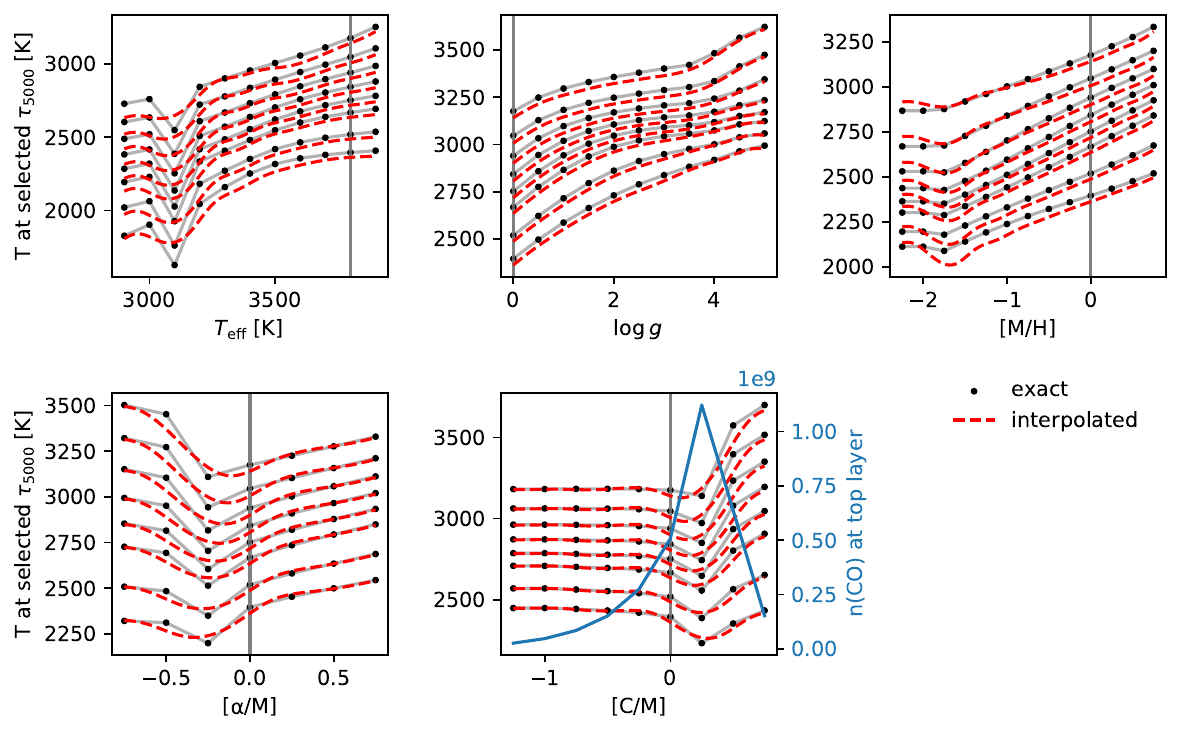}
    \caption{The temperature at selected upper layers in interpolated (using the resampled cubic method) model atmospheres ($10^{-4} \lesssim \tau_{5000} \lesssim 10^{-1}$) in exact (black) and interpolated (dashed red) model atmospheres.
    The interpolated atmospheres are always interpolated in all five stellar parameters.
    Each panel shows how the values change as one of the parameters varied and the others stay fixed.  
    The fixed values ($T_\mathrm{eff} = 3800~\mathrm{K}$, $\log g = 0.0$, solar abundances), are marked with vertical lines.
    In the last panel, the blue line indicates the number density of CO at the top of the model atmosphere, computed using the uninterpolated atmospheres only.
    Sharp changes in temperature as a function of stellar parameters driven by molecular formation make model atmosphere interpolation very challenging for cool giants.
    }
    \label{fig:itp}
\end{figure*}

Interpolating model atmospheres boils down to independently interpolating four (for cool dwarfs) or five depth-dependent quantities \emph{over} five stellar parameters.
Which of these quantities and parameters are driving errors in cool stars?
By replacing each depth-dependent quantity in a doubly interpolated atmosphere with the exact values, one at a time, we determined that the temperature at each layer is the largest driver of inaccuracy.
Figure \ref{fig:itp} demonstrates the effect of each stellar parameter on the interpolated temperatures at several layers.
By varying each stellar parameter in turn, and directly examining the interpolated temperatures, we can identify the parameters that would benefit from finer sampling.
While the temperatures at various optical depths are nearly linear as a function of $\log g$, and [M/H], they vary sharply as a function of $T_\mathrm{eff}$, [$\alpha$/M] and [C/M].
This is because of the sensitivity of molecular formation to these parameters.
For the case of [C/M], we also plot the number density of CO at the top of the atmosphere.
It's clear that the sharp temperature changes at $\mathrm{[C/M]} \approx 0.25$ is driven by opacity from CO.
We believe that other sharp transitions are similarly linked to formation of, and opacity from, specific molecules.

Finally, we note that the overwhelming challenges of model atmosphere interpolation for cool stars via standard techniques likely affects many analyses in the literature, as we are unaware of any past work documenting it.
As shown in Figure \ref{fig:itp}, finer sampling in most parameters is required to enable accurate interpolation in this regime.
In addition to denser sampling, additional abundance ratios will likely need to be included in model atmosphere grids to allow for accurate reconstruction of cool atmospheres.
As indicated in \citet{gustafssonGridMARCSModel2008}, nitrogen abundances should be included as a free parameter.
That work also indicates that the microturbulent velocity, which is especially impactful for cool, line blanketed stars, should also be allowed to vary.
\emph{Ad-hoc} calculation of model atmospheres may ultimately prove to be the only solution to fully address these problems.

\pagebreak
\section{internal improvements}\label{sec:internal}
\subsection{Equation of state and chemical equilibrium} \label{sec:eos}
The latest versions of \korg{} include substantial upgrades to the chemical equilibrium solver. 
\korg{} now self-consistently determines the electron number density, $n_e$, when solving for chemical equilibrium, rather than assuming the value provided by the model atmosphere.  
A configurable warning is raised when \korg{} produces a value substantially different from that in the model atmosphere.
The impact on the spectrum is nearly always undetectable, but allowing for $n_e$ to vary from the precise value in the model atmosphere allows \korg{}'s chemical equilibrium solver to converge for almost all stellar parameters.
Direct comparisons to the MARCS chemical equilibrium tables\footnote{Bengt Edvardsson, private communication} indicate that agreement is excellent except at the lowest temperatures, where small ($\lesssim 10^{-3}$) differences in Na and K ionization fraction drive moderate relative differences in the (very small) electron pressure.

In addition, \korg{} now supports polyatomic molecules, and includes partition functions for (the most abundant isotopologue of) each polyatomic molocule included in ExoMol\footnote{\url{https://www.exomol.com/}} version 20220926, except cis-P$_2$H$_2$ and trans-P$_2$H$_2$ (28 molecules total, listed in Table \ref{tab:mols}).
In order to compute equilibrium constants, atomization (dissociation) energies are required for each molecule.
These were calculated from the enthalpies of formation at 0~K of the molecule and its constituent atoms using experimental data from release 22 of the Computational Chemistry Comparison and Benchmark DataBase\footnote{\url{https://cccbdb.nist.gov/}}.
Conventionally, nuclear spin degeneracy is not included in molecular energy level degeneracies in stellar astrophysics, and care should be taken that all data, including linelist $\log gf$ values and partition functions use the same convention.
We obtained nuclear spin data from the Brookhaven National Laboratory ``Wallet Card'' service\footnote{\url{https://www.nndc.bnl.gov/nudat3/indx_sigma.jsp}} and used it to convert the ExoMol partition functions from the ``physics'' convention (including the nuclear spin degeneracy) to the ``astrophysics'' convention.

\begin{table}[]
    \centering
    \begin{tabular}{lp{6cm}}
        \textbf{Molecule} & \textbf{Reference} \\
        \hline 
        H$_{2}$O & \citet{polyanskyExoMolMolecularLine2018} \\
        CO$_{2}$ & \citet{yurchenkoExoMolLineLists2020} \\
        CH$_{4}$ & \citet{yurchenkoExoMolLineLists2014, yurchenkoHybridLineList2017} \\
        SO$_{2}$ & \citet{underwoodExoMolMolecularLine2016} \\
        NH$_{3}$ &  \citet{colesExoMolMolecularLine2019} \\
        HNO$_{3}$ & \citet{pavlyuchkoExoMolMolecularLine2015} \\
        H$_{2}$CO & \citet{al-refaieExoMolLineLists2015} \\
        HCN & \citet{harrisImprovedHCNHNC2006, barberExoMolLineLists2014} \\
        CH$_{3}$Cl & \citet{owensExoMolLineLists2018} \\
        H$_{2}$O$_{2}$ & \citet{al-refaieExoMolLineLists2016} \\
        C$_{2}$H$_{2}$ &  \citet{chubbExoMolMolecularLine2020} \\
        PH$_{3}$ & \citet{sousa-silvaExoMolLineLists2015} \\
        H$_{2}$S & \citet{azzamExoMolMolecularLine2016} \\
        C$_{2}$H$_{4}$ &  \citet{mantExoMolMolecularLine2018} \\
        SO$_{3}$ & \citet{underwoodExoMolMolecularLine2016a}  \\
        SiH$_{4}$ & \citet{owensExoMolLineLists2017} \\
        CH$_{3}$F &  \citet{owensRotationVibrationSpectrum2019} \\ 
        AsH$_{3}$ & \citet{colesVariationallyComputedRoom2019} \\
        PF$_{3}$ & \citet{mantInfraredSpectrumPF2020} \\
        CH$_{3}$ & \citet{adamVariationallyComputedIR2019} \\
        SiH$_{2}$ &\citet{clarkHightemperatureRotationvibrationSpectrum2020}  \\
        SiO$_{2}$ &\citet{owensExoMolLineLists2020}  \\
        KOH &  \citet{owensExoMolLineLists2021}\\
        NaOH & \citet{owensExoMolLineLists2021} \\
        CaOH & \citet{owensExoMolLineLists2022} \\
        H$_{2}$CS & \citet{yurchenkoExoMolLineLists2020}  \\ 
        \hline
   \end{tabular}
   \caption{Polyatomic molecules from the ExoMol project which have been added to \korg{}}
   \label{tab:mols}
\end{table}

Finally, \korg{} now includes support for positively charged molecules. 
By default, those with parition function in \citet{barklemPartitionFunctionsEquilibrium2016}: 
H$_{2}$+, He$_{2}$+, C$_{2}$+, N$_{2}$+, O$_{2}$+, Ne$_{2}$+, P$_{2}$+, S$_{2}$+, HeH+, BeH+, CH+, NH+, OH+, HF+, NeH+, MgH+, AlH+, SiH+, PH+, SH+, HCl+, ZnH+, HBr+, CdH+, HgH+, CN+, CO+, NO+, NS+, BO+, SiO+, PO+, SO+, AsO+, and TaO+.

\subsection{H I bf absorption and plasma effects}
\begin{figure} 
\centering
    \includegraphics[width=0.45\textwidth]{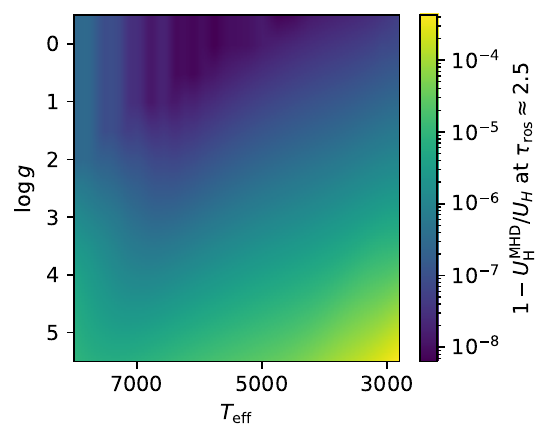}
    \caption{Fractional correction to the hydrogen partition function, $U_H$, deep in the stellar atmosphere according to the MHD formalism across the Kiel diagram.
    Stellar atmospheres are produced using \korg{}'s interpolation of the SDSS model atmosphere grid for solar-composition stars.
    Corrections are less than $4\times10^{-3}$ for all stars and much smaller than that at higher layers.}
    \label{fig:mhd_U_deviations}
\end{figure}

When atoms are embedded in a sufficiently hot and dense environment, their internal structure can no longer be dealt with in isolation.
The Mihalas--Hummer--Däppen (MHD) formalism \citep{hummerEquationStateStellar1988} provides a probability that each level is dissolved into the continuum, and thus a correction \revision{to} the partition function.
The formalism was initially developed to provide an equation of state for stellar interiors, but by changing level populations (of, e.g. hydrogen), it also predicts changes to the monochromatic opacity.
Plasma effects are typically subtle in stellar atmospheres, so the effect of MHD on the equation of state is negligible.
Figure \ref{fig:mhd_U_deviations} shows the fractional correction to the neutral atomic hydrogen (the species most strongly affected by plasma effects) partition function evaluated at $\tau_\mathrm{ros} = 2.5$, deep enough to have minimal impact on the spectrum.
The largest corrections are for the coolest main-sequence stars, but all corrections \revision{are} by factors of less than $4 \times 10^{-3}$.
Higher in the atmosphere, at depths that more strongly impact the emergent spectrum, the correction is smaller: $2\times10^{-4}$ at $\tau_\mathrm{ros} = 1$ and $3 \times 10^{-5}$ at $\tau_\mathrm{ros} = 10^{-2}$.
The corrections to level populations of hydrogen, however, are not negligible (see Figure 2 in \citealp{wheelerKORGModern1D2023}), which has implications for both bound-bound (line) and bound-free (photoionization) opacity.

\korg{} now includes the effects of the MHD formalism for hydrogen bound-free opacity, which has a particularly strong impact on the shape of the Balmer break.
Due to level dissolution, photons that would ordinarily excite the atom to an upper energy level (i.e. they have energy less than the classical ionization energy -- ${\sim}10.2~\mathrm{eV}$ from the $n=2$ level) have a finite probability of ionizing the atom instead.
This results in a ``rounded-off'' Balmer break.
This calculation requires knowledge of the ionization cross section for energies lower than the classical ionization energy.
We extrapolate the cross sections assuming that they are proportional to $E^{-3}$, where $E$ is the photon energy.
Using linear extrapolation instead results in differences in the cross section of up to a few percent.
Applying the MHD formalism to the Lyman-$\alpha$ transition results in additional unphysical absorption far into the visible for any reasonable method of extrapolating the photoionization cross section.
Because of this, we do not apply the MHD formalism to the Lyman series.

\citet{wheelerKORGModern1D2023} highlighted the fact that the application of MHD may over-attenuate hydrogen lines (see Section \ref{sec:brackett} for a related discussion).
It has been noted previously that MHD dissolves outer, sparsely populated orbitals more eagerly that the OPAL \citep{rogersOPALEquationofStateTables1996a} equation of state (e.g. \citealp{iglesiasDiscrepanciesOPALOP1995}).
\citet{nayfonovMHDEquationState1999a} presents a modified version of the MHD formalism that may address this in the stellar atmosphere regime, but further investigation is required.
While we include the effects of MHD on hydrogen bound-bound opacity by default, we provide a switch to turn it off if desired.

\subsection{Brackett lines: plasma effects and broadening mechanisms in the infrared}\label{sec:brackett}
Spectral synthesis codes must treat hydrogen lines specially (if they are to model them accurately), because they are subject to resonant and linear Stark broadening.
Support for Brackett series lines has been added to \korg{}, extending its applicability further into the infrared (out to about $2.3~\mathrm{\mu m}$, the Pfund series limit).
\korg{}'s implementation is adapted from the \codename{HLINOP} hydrogen opacity code \citep{barklemHydrogenBalmerLines2003, barklemHlinopApril20162016} and translated into \codename{Julia} to facilitate automatic differentiation.
\codename{HLINOP} handles Brackett lines using the theory developed in \citet{griemStarkBroadeningHigher1960}, which extended the classic Holtsmark theory \citep{holtsmarkUberVerbreiterungSpektrallinien1919} to include the effects of ions, and \citet{griemCorrectionsAsymptoticHoltsmark1967}, which added a correction for distant electrons.
Comparisons between \codename{Synspec} \citep{hubenySynspecGeneralSpectrum2011,hubenyBriefIntroductoryGuide2017,hubenyTLUSTYSYNSPECUsers2021} and \codename{Turbospectrum} \citep{plezLithiumAbundancesOther1993,plezTurbospectrumCodeSpectral2012,gerberNonLTERadiativeTransfer2022} (used to generate model spectra for APOGEE in DR 17, \citealp{abdurroufSeventeenthDataRelease2022}) revealed that agreement on the Brackett lines is particularly poor, in contrast with other lines in the infrared, where synthesis codes tend to agree well \citep{wheelerKORGModern1D2023}.
Since \korg{}'s implementation was heavily inspired by \codename{HLINOP}, the initial implementation produced Brackett lines which matched \codename{Turbospectrum}'s almost exactly, but the lines produced by \texttt{Synspec} differed significantly.
Upon investigation, two independent factors were determined to be driving the disagreement between models: treatment of level dissolution, and oversights regarding line broadening mechanisms in the infrared.
The first factor is simple: \codename{Turbospectrum} includes the effects of the MHD formalism on hydrogen lines, while \codename{Synspec} seems not to.
The second factor requires slightly more explanation.

\begin{figure}
    \centering
    \includegraphics[width=3.0in]{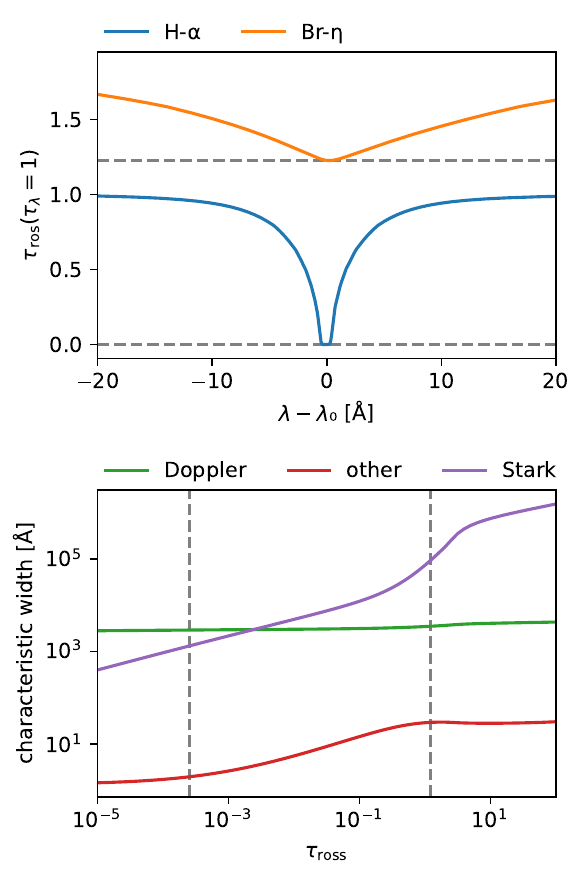}
    \caption{\textbf{Top:} the Rosseland mean optical depth of formation of \halpha{} and \breta{} as a function of wavelength. Lines in the infrared form much deeper in the stellar atmosphere, where different broadening mechanisms dominate.
    \textbf{Bottom:} The characteristic profile widths of line broadening mechanisms as a function of depth. The dashed lines in each panel mark the Rosseland mean optical depth at which the core of each line forms. For line cores in the visible, which form high in the atmosphere, Doppler broadening dominates, but for line cores in the infrared, which form deep in the atmosphere, Stark broadening dominates.
    These calculations were done using a model atmosphere with $T_\mathrm{eff} = 6000 ~\mathrm{K}$, $\log g = 3.5$.}
    \label{fig:formation_depth}
\end{figure}

The top panel of Figure \ref{fig:formation_depth} shows the depth of formation for two hydrogen lines: \halpha{} (visible), and \breta{} (infrared).
The core of \halpha{} forms high in the atmosphere, at a Rosseland mean optical depth of $10^{-3.5}$ or so.
Because the monochomatic opacity is lower in the infrared than in the visible, the core of hydrogen lines in the infrared form much deeper, typically below the Roseland mean photosphere.
This contrast has important implications for line broadening mechanisms---hydrogen lines that form deep in the atmosphere don't have Doppler-dominated cores.
The second panel of Figure \ref{fig:formation_depth} shows the characteristic widths of hydrogen line profiles from the Doppler broadening, linear Stark broadening, and from other mechanisms (fundamental line widths and van der Waals broadening).
At the depth of formation of \halpha{} (marked with a dashed line), the Doppler broadening profile is the widest, so it sets the shape of the absorption profile resulting from convolving all three profiles.
But at the depth of formation of \breta{}, Stark broadening sets the shape of the line core.
While none of the above constitutes novel theoretical insight, synthesis codes developed for application to the visible and near ultraviolet have generally overlooked the fact that hydrogen lines in the infrared need separate consideration.

The reason for the oversight is that numerically convolving broadening profiles is computationally expensive (in decades past, prohibitively so).
As a consequence, all spectral synthesis codes commonly used for fitting data employ a variant of the same approximation: the line wings are obtained by adding all profiles, and the line core is set to the broadest one.  
This approximation works remarkably well, though problems arise at the boundary between the core and the wings, and when two of the broadening profiles have similar widths (this case arises for Brackett lines in cool stars, where the Stark and Doppler widths at the depth of formation can be comparable).
For some codes, the issue is that the core is hard coded to be Doppler-dominated.
For codes whose hydrogen lines descend from the Kurucz routines, 
the issue is more subtle.
The \citet{griemStarkBroadeningHigher1960, griemCorrectionsAsymptoticHoltsmark1967} theory employed provides separate profiles for Stark broadening in the quasistatic and impact limits. 
Profiles for each must be calculated using the number of perturbers (approximately) in each regime, then the profiles are convolved to give the total Stark profile.
These codes add the two profiles together, which gives a good approximation to the profiles in the wings, but results in a core with far too much opacity, and wavelength-integrated cross sections which are wrong by a factor of $\sim 2$ when Stark broadening is dominant.
Thankfully, this problem is easily remedied by using true numerical convolution of the broadening profiles.

\begin{figure*} 
    \centering
    \includegraphics[width=\textwidth]{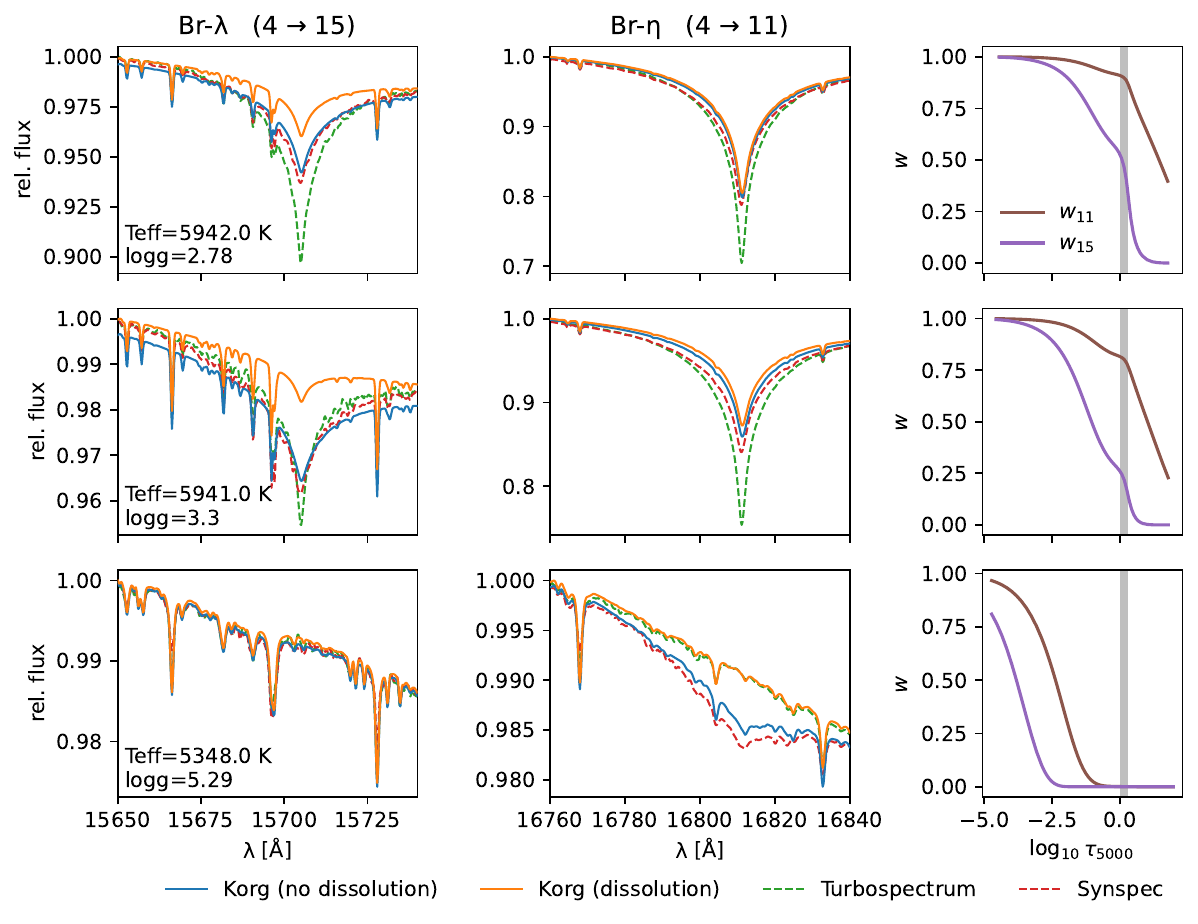}
    \caption{Predicted lines of \breta{} and \brlamda{} ($n=4$ to $n=11$ and $n=15$, respectively), two lines in the \apogee{} wavelength range.
    Agreement between synthesis codes is poor because models don't properly account for line formation in the infrared and because of differing treatments of level dissolution. 
    Each row represents a different set of stellar parameters ($T_\mathrm{eff}$ and $\log g$ are indicated in the first column, and all models are computed with $\mathrm{[M/H]} = -1.5$).
    The third column shows the probability that the $n=11$ and $n=15$ orbitals survive (are not dissolved into the continuum) as a function of optical depth at $5000~\mathrm{\AA}$, $\tau_{5000}$. 
    The vertical gray band marks the approximate depth of formation of the Brackett lines.
    These spectra were synthesized using the APOGEE DR 17 linelist and convolved to the instrument resolution ($R = 22,500$).
    }
    \label{fig:brackett_theory}
\end{figure*}

Figure \ref{fig:brackett_theory} illustrates the effects of level dissolution and of broadening treatment on the Brackett series.
It shows Br--$\eta$ (the strongest hydrogen line in the APOGEE wavelength range), and Br--$\lambda$ (a weaker line) for a few stellar parameters, as predicted by \codename{Synspec}, \codename{Turbospectrum}, and \korg{}.
For each star, the probability that the upper level of each transition is \emph{not} dissolved into the continuum, $w$, is plotted as a function of optical depth, with a gray band marking the depth of formation for the Brackett series.
In some cases, MHD results in highly attenuated lines.
Contrast \codename{Synspec} and \korg{} without dissolution to \codename{Turbospectrum} and \korg{} with dissolution in Br--$\lambda$ in row 2 and in Br--$\eta$ in row three.
When level dissolution is weak (both lines in row 1, Br--$\eta$ in row 2), the differences in predicted line profiles are due to differences in the broadening of the line core.

\begin{figure*} \label{fig:brackett_benchmarks}
    \centering
    \includegraphics[width=\textwidth]{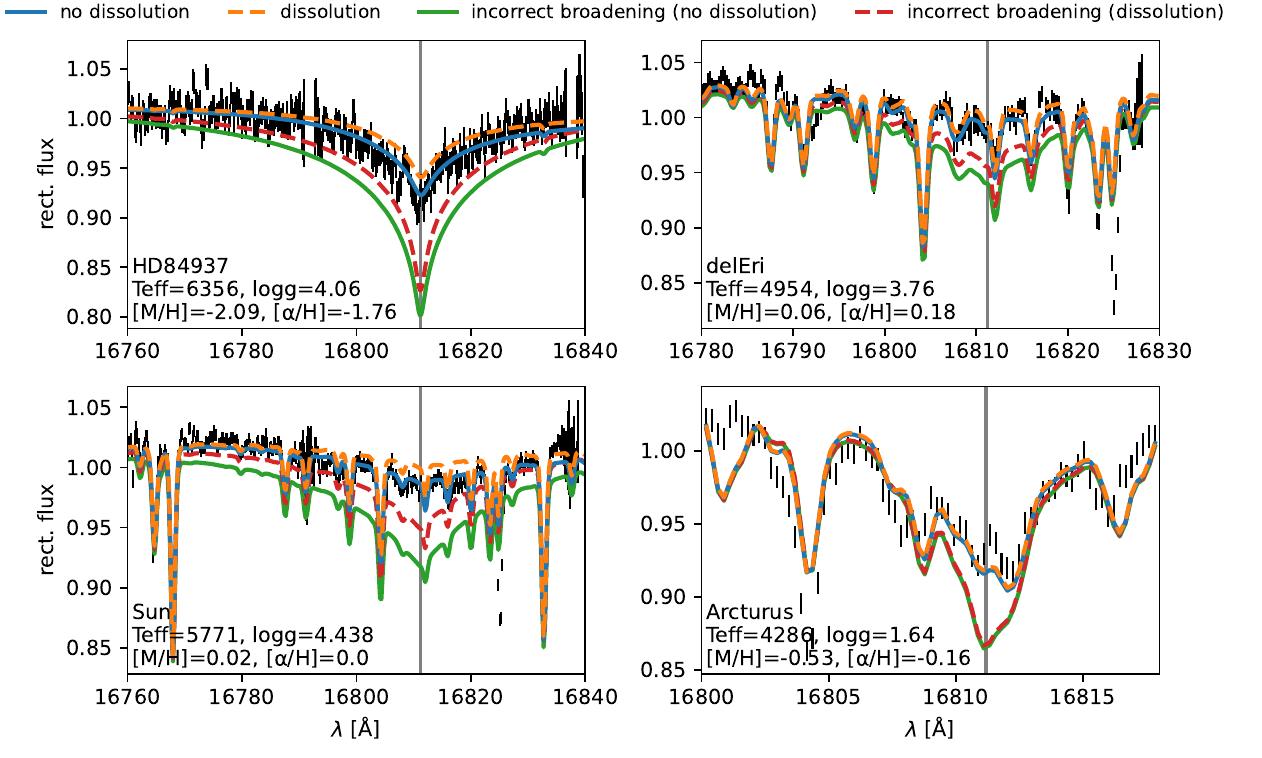}
    \caption{
    Comparisons of synthesized Br--$\eta$ profiles for four representative Gaia FGK benchmark stars. 
    The fit is much improved using the corrected profiles, though the effect of the MHD dissolution formalism is to subtle to constrain using these data.
    }
\end{figure*}

Figure \ref{fig:brackett_benchmarks} compares our improved Brackett profiles to a representative selection of Gaia FGK benchmark stars \citep{blanco-cuaresmaGaiaFGKBenchmark2014} observed by APOGEE.
We show the Br--$\eta$ lines computed using the naive and corrected profiles, and with level dissolution switched on and off.
The lines are synthesized using the parameters from \citet{blanco-cuaresmaGaiaFGKBenchmark2014}, including nonspectroscopic $T_\mathrm{eff}$ and $\log g$.
These comparisons clearly indicate that the corrected profiles significantly improve the accuracy of the synthesized spectra.
Unfortunately, none of the benchmark stars provide a clear constraint on the effects of level dissolution, since including it doesn't consistently improve or degrade the fit.
Comparisons for all Gaia FGK benchmark stars observed for APOGEE DR 17 are in Appendix \ref{sec:brackett_appendix}.

Finally, we note that ideally stellar spectral synthesis codes would treat the Brackett series using the model microfield method \citep{brissaudTheoryStarkBroadening1971}, as is often done for other hydrogen series via the tables of \citet{stehleExtensiveTabulationsStark1999}.
In lieu of this, Vidal--Cooper--Smith (VCS) theory \citep{smithUnifiedClassicalPathTreatment1969, voslamberUnifiedModelStark1969, vidalHydrogenStarkBroadening1970, vidalUnifiedTheoryCalculations1971} as calculated by \citet{lemkeExtendedVCSStark1997} may be a promising alternative.

\subsection{Linelist parsing}
In addition to Vienna Atomic Line Database\footnote{\url{http://vald.astro.uu.se/}} (VALD) linelists, \korg{} now supports a \texttt{MOOG} linelists with isotope information, ``Kurucz''-format molecular linelists, and \texttt{Turbospectrum} linelists.
When parsing linelists with isostopic information, \korg{} adjusts the $\log gf$ value of each line to account for isotopic abundances, which can be specified by the user (except in the case of pre-adjusted VALD linelists).

\section{Conclusions} \label{sec:conclusions}
We have presented several updates to the \korg{} code for stellar \revision{spectral} synthesis.
\revision{The version of the code used in this work is 0.27.1}.
The new capabilities include routines for fitting observed spectra via either synthesis or equivalent widths.
The equivalent width fitting routine was tested against \citet{melendez18ScoSolar2014} and found to produce iron abundances with a significantly lower scatter than the original analysis.

We have described the model atmosphere interpolation method employed by \korg{} and shown that it introduces minimal error to synthesize spectra for stars with $T_\mathrm{eff} > 4000\,\mathrm{K}$.
Stars with $T_\mathrm{eff} < 4000\,\mathrm{K}$ have atmospheres that contain molecules in significant quantities, which greatly \revision{complicate the dependence} of atmospheric thermodynamic quantities on stellar parameters.
Flux errors introduced by interpolation are smaller in the infrared than the visible, because lines in the infrared form at deeper atmospheric layers where thermodynamic quantities are better interpolated.
While it it possible to obtain sub-percent atmosphere interpolation error for cool dwarfs, finer sampling in stellar parameters is required if we are to make interpolation error truly negligible.
Additionally, more parameters and abundances than are typically taken into account (e.g. microturbulence) are likely to impact molecular densities and opacities.
It remains to be seen whether sufficiently-dense grids of cool star model atmospheres are feasible to generate and use; \emph{ad-hoc} model atmosphere calculations may be necessary.

\korg{}'s chemical equilibrium solver has been significantly updated to facilitate synthesis for cool stars.
Support for charged and polyatomic molecules has been added, with state-of-the-art partition functions from the ExoMol project.
The electron number density is now calculated when solving the chemical equilibrium equations, meaning that syntheses with abundances varying from the model atmosphere are more internally consistent.

We now apply the Mihalas--Hummer--Däppen formalism to neutral atomic hydrogen, allowing the outer orbitals of hydrogen to dissolve into the continuum in dense regimes.
While this treatment (or something similar) is both physically motivated and necessary to reproduce a ``rounded-off'' Balmer break, there are shortcomings to the theory applied in the stellar atmosphere regime.

Finally, we highlight the fact that synthesis codes produce Brackett series lines that are in strong disagreement with each other.
We show that the causes of the disparity are the treatment of level dissolution and that the fact that Stark broadening dominates hydrogen line \revision{cores} in the infrared has been frequently overlooked.
Comparisons to spectra of stars with non-spectroscopic values of $T_\mathrm{eff}$ and $\log g$ indicate that our corrected Brackett profiles are in much better agreement with the data than those of other codes.

\section*{Acknowledgements}

We thank 
 Paul Barklem (Uppsala) for helpful correspondence,
 Bengt Edvardsson (Uppsala) for sharing the MARCS equation of state data,
 Emily Griffith (CU) for providing observational data,
 Jon Holtzman (NMSU) for suggesting the method of interpolating grids halfway between points, then interpolating back to the original points,
 Sergey Koposov (Edinburgh) for motivating the investigation into H I bound-free absorption and plasma effects, and
 Andrew Saydjari (Harvard) for extensive stress testing of the code.
 
\bibliography{Korg2}

\appendix

\section{Brackett lines in benchmark stars} \label{sec:brackett_appendix}
\begin{figure}
    \centering
    \includegraphics[width=\textwidth]{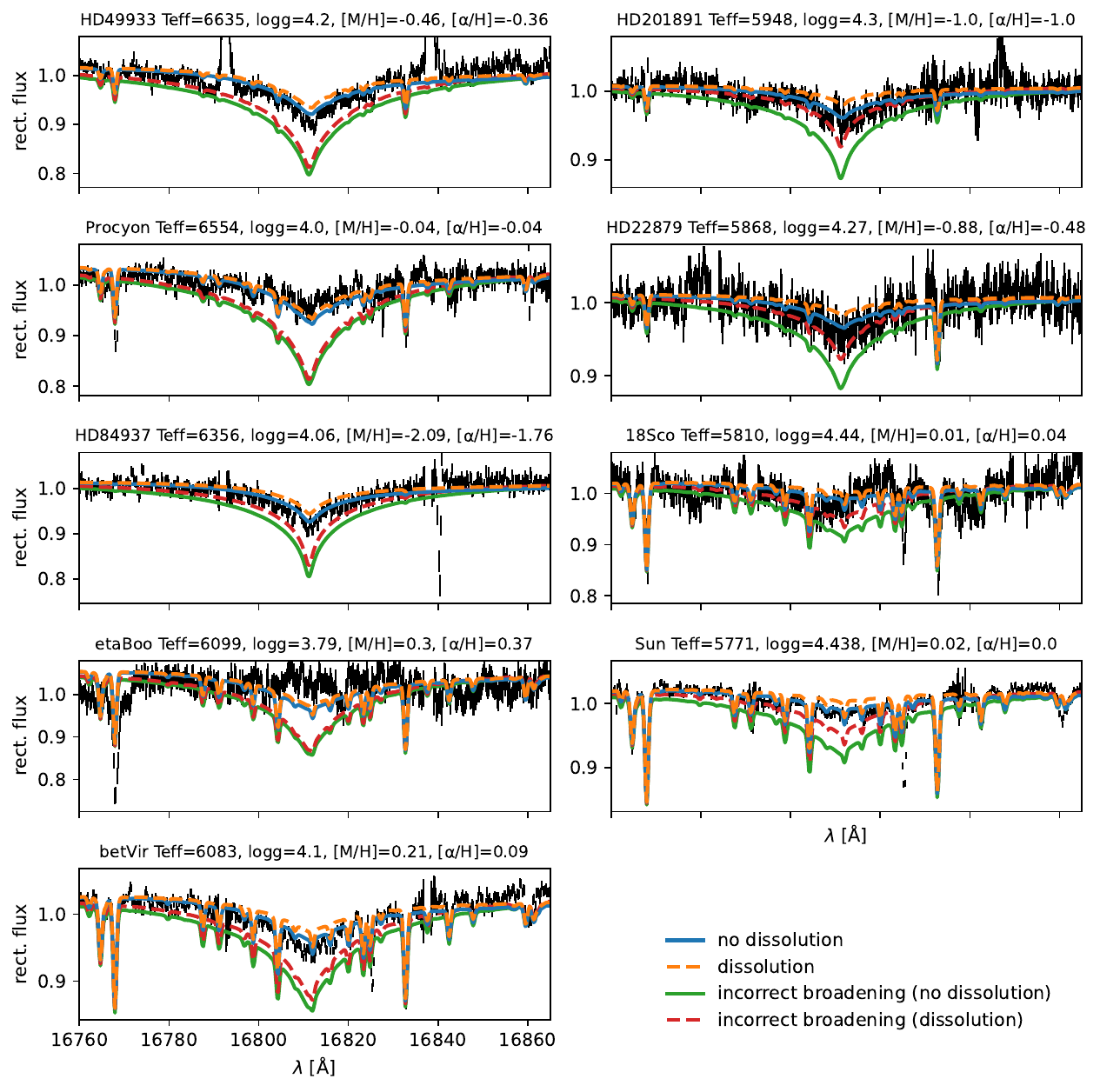}
    \caption{Br-$\eta$ in Gaia FGK benchmark stars with $T_\mathrm{eff} > 5500~\mathrm{K}$ compared to \korg{} spectra using the legacy and corrected Brackett line profiles with level dissolution included and neglected. 
    The updated profiles are a much closer match to the spectra in nearly all cases.
    }
    \label{fig:benchmarks_brackett_lines_1}
\end{figure}

\begin{figure}
    \centering
    \includegraphics[width=\textwidth]{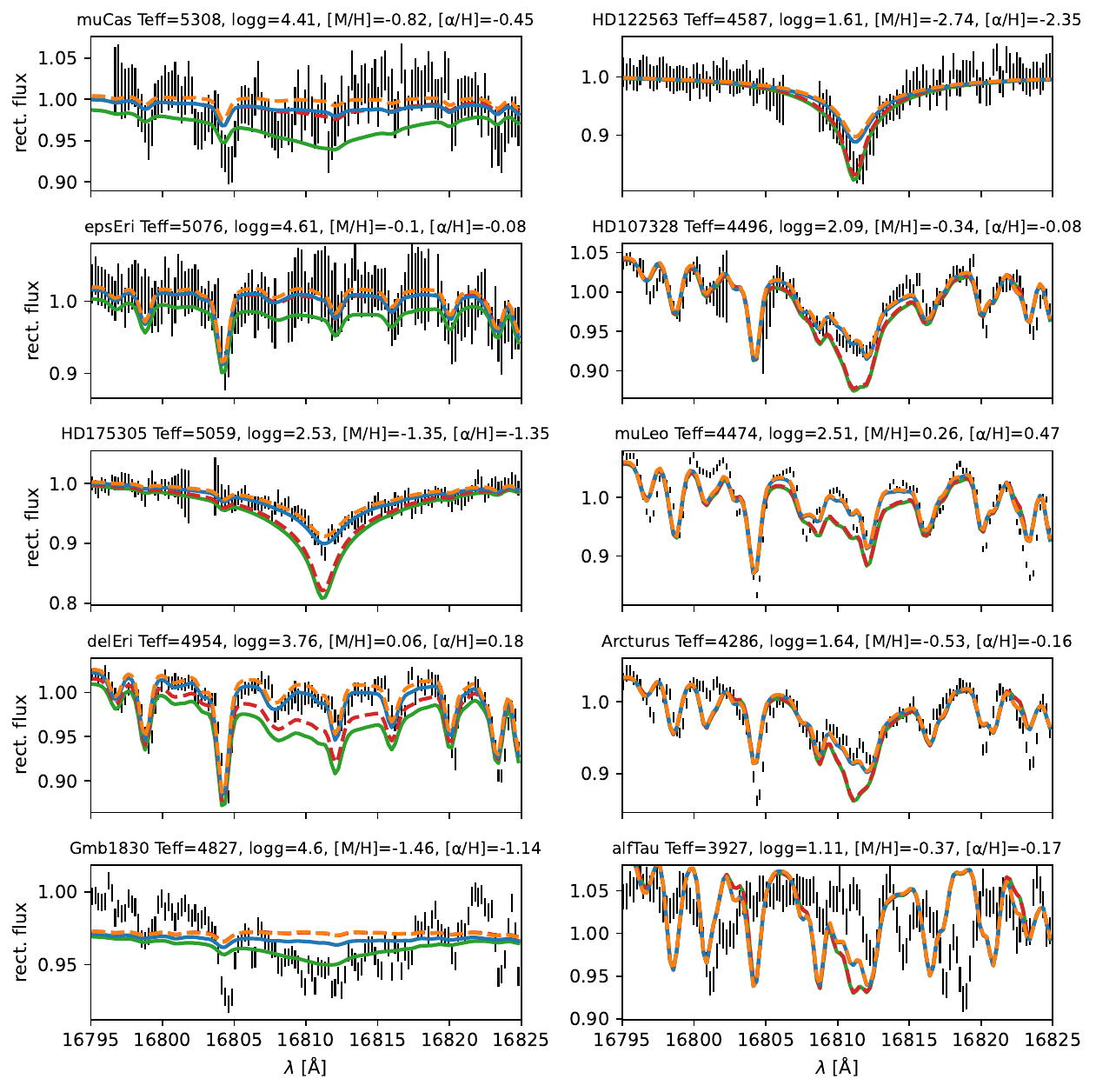}
    \caption{
    The same as Figure \ref{fig:benchmarks_brackett_lines_1} for stars with $T_\mathrm{eff} < 5500~\mathrm{K}$.
    }
    \label{fig:benchmarks_brackett_lines_2}
\end{figure}

Figures \ref{fig:benchmarks_brackett_lines_1} and \ref{fig:benchmarks_brackett_lines_2} show Br--$\eta$ for each Gaia FGK benchmark star which is part of APOGEE for DR17.
For all stars, the benchmark parameters were used for \revision{spectral} synthesis, treating [Fe/H] as a proxy for [M/H] and [Mg/H] as a proxy for [$\alpha$/H].
For most stars in which Br--$\eta$ can be observed, the fit with the corrected Brackett profiles is excellent.
Comparisons to Br--$\lambda$ are less illuminating since the line is weaker and none of the benchmark stars have parameters such that the combination of the two lines provides a good test of the MHD dissolution theory.
Eta Boo, Gmb1830, HD 122563, and Alpha Tau are the stars for which the match is the worst.
For Eta Boo, we speculate that the abundances may be substantially wrong, or that there is a problem with the APOGEE DR 17 linelist.
We have verified that the radial velocity shift appears to be correct.
The problems with the other three stars may arise from errors in the stellar parameters or from missing physics in the model atmospheres.


\end{document}